\documentclass[a4paper]{article}
\usepackage{lmodern}
\usepackage[T1]{fontenc}
\usepackage[utf8]{inputenc}
\usepackage{amssymb,amsmath}
\usepackage{graphicx}
\usepackage{xcolor}
\usepackage{soul}

\usepackage{natbib}
\usepackage{hyperref}
\hypersetup{unicode=true,
            pdfborder={0 0 0},
            breaklinks=true}
\usepackage{lineno}

\usepackage{longtable,booktabs}
\IfFileExists{parskip.sty}{%
\usepackage{parskip}
}{
\setlength{\parindent}{0pt}
\setlength{\parskip}{6pt plus 2pt minus 1pt}
}
\title{Reconfigurable mechanical vibrations laboratory kit}
\usepackage{authblk} 
\author[1]{Ian Ho}
\author[1]{Ajay Harishankar Kumar} 
\author[1*]{Daniel M. Harris} 
\affil[1]{Brown University, School of Engineering, 184 Hope St., Providence RI 02912} 
\affil[*]{Correspondence email address: daniel\_harris3@brown.edu}

\begin{document}
\maketitle

\begin{abstract}
Mechanical vibrations appear across a diverse range of applications, but nevertheless are unified by a core set of principles and phenomena.  In this article, we describe a reconfigurable laboratory kit that is designed to be used as part of a first undergraduate course in mechanical vibrations.  The kit was designed to be low-cost and allow for a series of laboratory experiments to be performed remotely.  Using a common set of core hardware and software, five laboratory experiments are detailed and can be performed to supplement many of the major topics traditionally encountered in a first vibrations course.  This accessible platform will readily enable further technical improvements and expansions, as well as the development of additional experiments and assignments.

\end{abstract}

\begin{longtable}[]{@{}l@{}}
\begin{minipage}[t]{0.97\columnwidth}\raggedright\strut

\subsection{Metadata Overview}\label{h.akaipbqoqfs8}

Main design files: {\color{black} \href{https://doi.org/10.5281/zenodo.5888554}{https://doi.org/10.5281/zenodo.5888554}}

Target group: undergraduate students in engineering, physics.

Skills required: 3D printing - easy; laser cutting - easy; metal shear and hole punch - easy.

Replication: Kit has been constructed and tested by authors, as well as students at Brown University. 

See section ``Build Details'' for more detail.

\subsection{Keywords}\label{h.kdz351yp7g7c}

{~open-source; vibration; resonance; oscillation; spring; laboratory; remote learning; teaching; mechanical engineering}

\strut\end{minipage}\tabularnewline
\bottomrule
\end{longtable}

{\color{black}\section{(1) Overview}\label{h.f8237gmzmwc6}}
\subsection{Introduction}\label{h.pnj38xyr5dyy}

Mechanical vibrations and associated phenomena are omnipresent to the world around us, from quantum \citep{nielsen1951vibration} to oceanic scales \citep{garrett1979internal}.  For engineering applications in particular, mechanical vibrations can either be undesirable (e.g. leading to the well-known failure of the Tacoma Narrows Bridge \citep{billah1991resonance}) or desirable (e.g. for applications in energy harvesting \citep{wei2017comprehensive}).  Due to their broad relevance, mechanical vibrations are almost universally taught at the university level as part of the core mechanical engineering curriculum.
Regardless of the ultimate application, mechanical vibrations are connected by a common thread of underlying principles and phenomena.  In this work, we present a low-cost reconfigurable laboratory kit which visually showcases a number of the principal phenomena encountered within a first undergraduate course on mechanical vibrations.  The open-source nature of the hardware will facilitate design of additional modules to supplement the five laboratory experiments already developed and described herein.

The hardware was designed in the midst of the COVID-19 pandemic, which forced nearly all university students into remote learning environments.  While challenging in myriad ways, this abrupt change in our learning environment also laid fertile ground for critical reflection and inspired pedagogical innovations that may persist beyond the pandemic.  Our goal with the development of the laboratory kit was to replace a traditional shared (and instructor maintained) laboratory equipment in a junior/senior-level elective entitled `Vibration of Mechanical Systems' with a series of physical experiments that could be completed entirely remotely.  These experiments paralleled the theoretical and computational framework introduced via traditional (in this case remote) lectures.  Due to the number of kits that needed to be developed with limited resources, attention was paid to overall cost and simplicity during the design and fabrication of the laboratory kit. {\color{black} Similar educational-grade demonstration platforms are available commercially, however they are significantly more expensive{\color{red},} which in our case was prohibitive to the distribution of an individual kit to each student.  For instance, Pasco (pasco.com) sells a Mechanical Wave Driver and Sine Wave Generator which total over \$500 USD \-- an order of magnitude more costly than the DIY shaker system described herein.} The heart of our kit is a small loudspeaker, which serves as the driver of forced periodic vibrations. Numerous reconfigurable appendages can be attached to the driver, and share a common set of parts.  In Fall 2020, the laboratory kit was mailed (or distributed locally) to all student participants worldwide who were enrolled in the course at Brown University.  All fifteen students successfully completed the five laboratory experiments and assignments.

In this work, we describe the overall implementation and design of the laboratory kit.  Five labs are then described in detail, with representative data (in the form of visualizations) accompanying the descriptions.  We conclude by discussing and summarizing student feedback as well as potential future directions for the work.  All source files and documentation are made available with the manuscript to allow for recreation, redesign, and expansion of the complete device and associated labs. 

\subsection{Overall Implementation and design}\label{h.1u7vph94gfbt}
The hardware of the vibrations laboratory kit contains two parts: a housing assembly and an electronics kit. The housing assembly consists of rectangular panels which are laser cut (Universal Laser Systems, VLS 4.60 with 2.0 focusing lens) out of 1/4" medium density fiberboard (MDF) using a 30 Watt CO$_2$ laser. The housing panels are designed with notches to allow them to be joined easily. The panels contain insert slots that allow fasteners to be embedded in the assembly to secure the structure. As a result, the housing can easily be deconstructed and reconstructed without the use of adhesives. The housing assembly also contains mounting through holes to fasten a shaker (loudspeaker) to the housing. 

\begin{figure}[hbt!]
    {\centering
    \includegraphics[width=350pt]{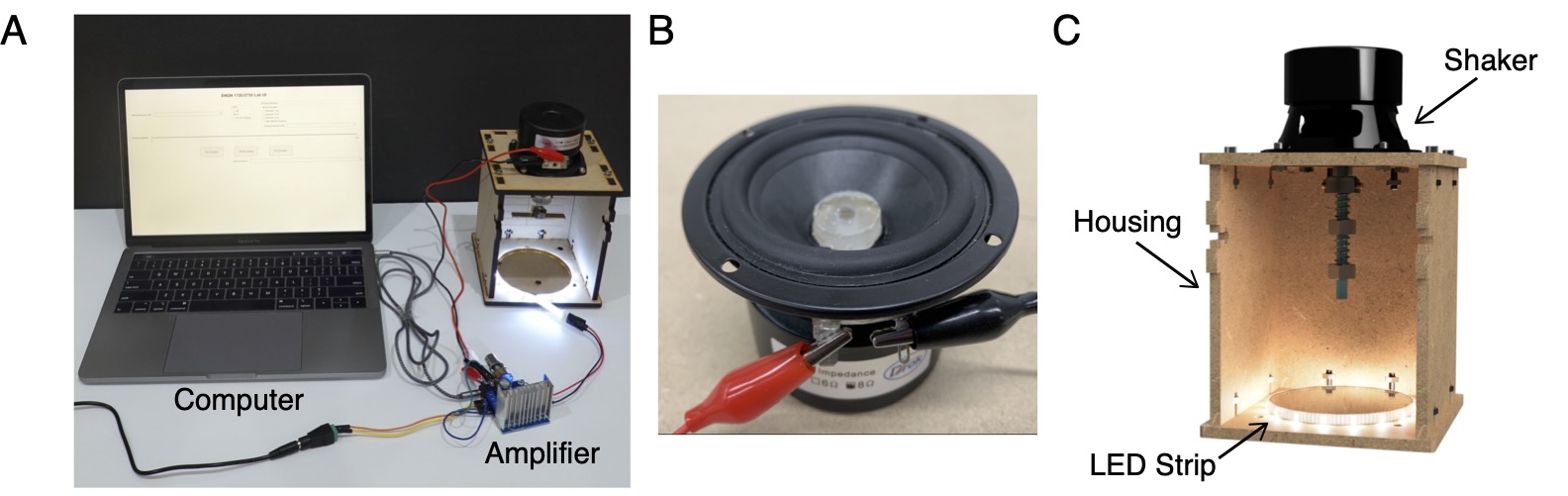}
    \caption{\small Vibrations laboratory kit. (A) Image of the experimental setup including the housing, electronics, and the software (GUI) being operated on a computer. {\color{black} (B) Close-up image of speaker with cylindrical mount epoxied to voice coil}. (C) CAD rendering of the housing assembly including the vibration driver and LED strobe lighting. }
    }
    \label{fig:setup}
\end{figure}

\begin{figure}[hbt!]
    {\centering
    \fbox{\includegraphics[width=350pt]{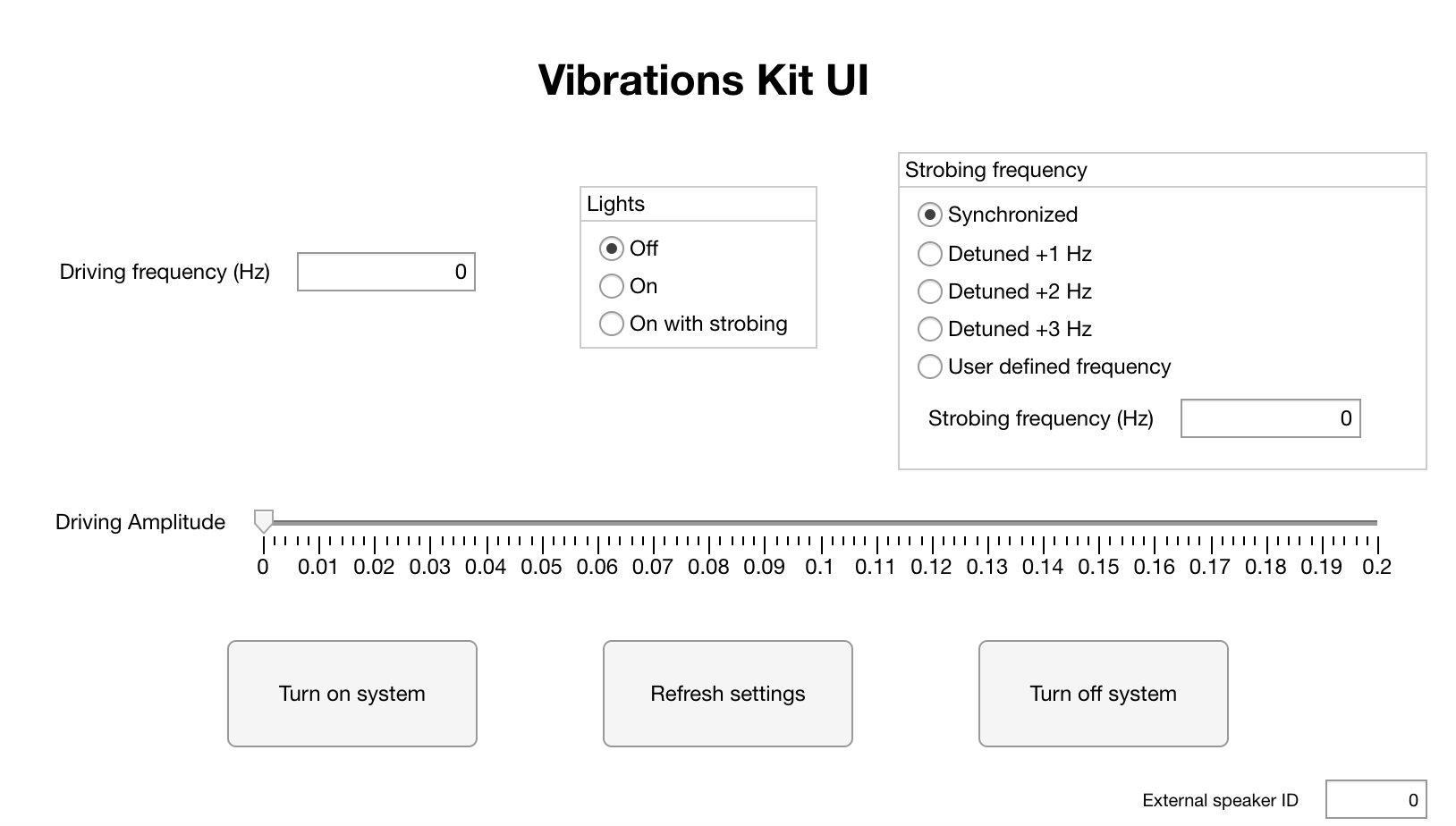}}
    \caption{\small Graphical user interface (GUI) developed and operated in MATLAB to control the vibration and lighting in the laboratory kit.}
    }
    \label{fig:GUI}
\end{figure}

\begin{figure}[hbt!]
    {\centering
    \includegraphics[width=350pt]{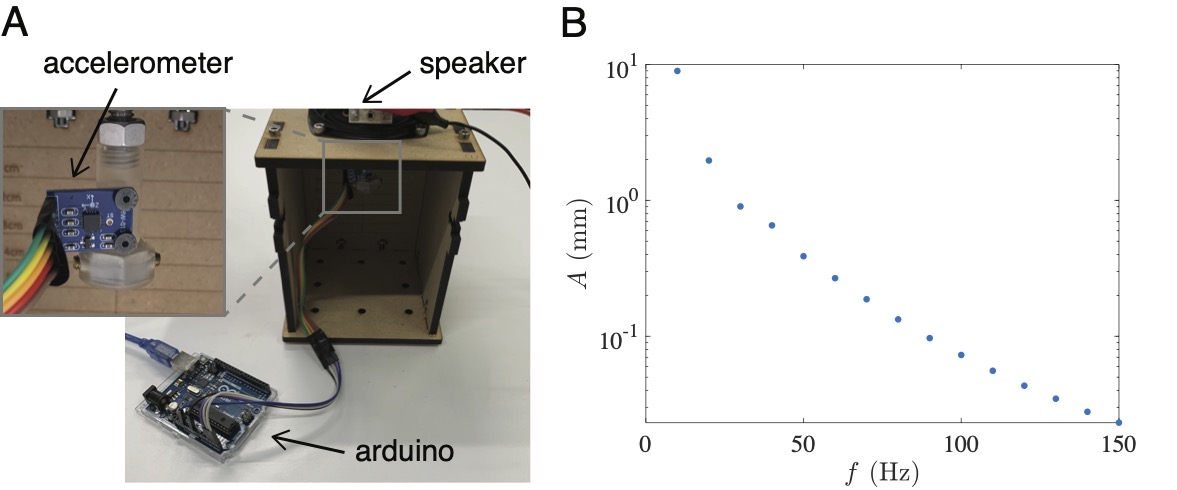}
    \caption{\small (A) Shaker characterization test setup. (B) Maximum vibration amplitude versus frequency for DIY shaker.  Peak accelerations $\gamma$ were converted to equivalent sinusoidal amplitudes $A$ using the formula $A={\gamma}/{(2\pi f)^2}$.}
    }
    \label{fig:shaker_test}
\end{figure}

The electronics assembly is centered around a DIY electrodynamic shaker. Shakers may be available in some engineering classrooms but are generally expensive (e.g. Fisher Scientific educational vibration generator, \$130). For our laboratory kit, we modified a 3", 15 Watt speaker (\$18) for this purpose: the dust cap of the speaker is removed, and a 3D printed (Form 2, clear resin, 0.5mm resolution) threaded cylinder is epoxied to the inner surface of the voice coil.  The attachment allows a drive rod to be threaded to the speaker. While effective for many applications, this DIY shaker has limitations. In particular, the amplitude of the small speaker is limited by both power and mechanical constraints (for the latter, {\color{black} if the oscillation amplitude is too high, the attached cylinder can physically contact the speaker's underlying magnet on each cycle, which we refer to as ``bottoming out''}).  For the case of bottoming out, we observed that as long as the amplitude is reduced whenever this happens, no permanent damage is done to the speaker. Despite such drawbacks, the speaker was nonetheless sufficient for the purposes of the experiments. The speaker is connected to a dual-channel digital amplifier powered by a 12V power source. The amplifier has an AUX cable port so that the speaker can be driven by a computer using a custom code and graphical interface written in MATLAB. The second output of the amplifier is connected to a set of LED lights that provides additional illumination and can be set to strobe in sync or out of sync with the speaker's motion for enhanced visualization. The electronics kit can be assembled easily using commercially available parts. Furthermore, all electronic connections use jumper wires with either ``screw to connect'' or ``alligator clip'' connections to allow for both increased safety and ease-of-use during assembly. Figure \ref{fig:setup}1A illustrates the entire setup, including the aforementioned electronics and hardware assemblies, {\color{black} with the shaker  {and mount shown separately in Figure \ref{fig:setup}1B}. Figure \ref{fig:setup}1C shows a CAD rendering of the housing assembly. 

In order to drive the shaker (via the AUX cable), we developed a graphic user interface (GUI) in MATLAB. The GUI consists of a text box for the user to enter the desired frequency, a slider bar to adjust the relative amplitude of the shaker, and radio buttons to trigger the lights between being in the on, off, or various strobing states. The GUI can be used immediately once the MATLAB script is run, but requires that the user first identify the external speaker ID number of the connected AUX cable with a simple command. The layout of the GUI is shown in Figure \ref{fig:GUI}2.  {\color{black} A simple but fully functional alternative to the MATLAB GUI has also been developed in Octave.}

The hardware and software described constitute the core components of the laboratory kit. With additional hardware, the housing assembly can be modified to accommodate five different demonstrations. Certain custom parts require the use of a 3D printer or laser cutter. In some of these components, threading is required {\color{red},} and a hand tap was used to re-thread printed components or thread through holes in laser cut components.

\section{(2) Quality control}\label{h.f8237gmzmwc6}

\subsection{Safety}\label{h.v60aduckfisj}
There are very few hazards present when assembling and operating the device, although we describe important safety considerations in what follows.  While operating the lab kit, all liquids should be kept away from the electronics in the system.  It is recommended that the power source for the electronics remain unplugged during construction, reconfiguration, or when not in use.  The lab kit also incorporates strobed LED lights to aid visualization of the various vibration phenomena.  The use of strobing is strictly optional and should be avoided if the user has any known sensitivity to flashing lights. In one of the use cases, a {\color{black} ferromagnetic} sheet metal beam is required which may have sharp edges. As a result, special care should be taken while handling sheet metal, and all edges and corners should be filed and deburred before use. 


{\color{black}\subsection{Startup procedure}\label{h.kr90wh14sxr5}}
A brief startup procedure is required to properly connect the shaker and tune the amplitude of the shaker's vibration as well as that of the LED lights. Once the shaker is connected to a computer and the MATLAB script has been loaded, the port ID of the AUX cable must be manually identified in order for the GUI to interface with the shaker properly. In order to do this, the command {\color{black} \texttt{info = audiodevinfo}} must be run, the output port's ID navigated to from MATLAB's workspace, and the ID input into a text box within the GUI front panel. There are two means to tune the amplitude of the output signals to the shaker: a slider bar that is included in the GUI and the computer's output volume. The  LED lights require a minimum signal threshold strength to illuminate and their brightness depends on the computer's output volume. In order to ensure that both the shaker and the LED lights can be triggered simultaneously, the driving amplitude slider bar should first be set to zero amplitude, while the volume of the computer should be slowly increased from zero until the LED lights are visible (with the lights in the ``on'' option selected in the GUI). Once this occurs, the computer's output volume should be fixed for the duration of use. Next, the shaker should begin to vibrate at the desired frequency after slowly increasing the amplitude displayed on the slider bar. Beyond a certain amplitude, the shaker's cylindrical mount may clash with the underlying magnet; if this happens, the amplitude should be decreased and maintained below this threshold using the slider bar henceforth.  This maximum amplitude depends on the driving frequency, payload, and orientation of the drive axis relative to gravity. {\color{black} An equivalent startup procedure for the Octave alternative is detailed in the Assembly Guide.}




\subsection{General testing}\label{h.wbekh9ay82yu}
In order to characterize the performance of the DIY shaker, an analog accelerometer (ADXL335 - 5V triple-axis accelerometer) was interfaced with an Arduino Uno and mounted to the drive rod via a custom 3D printed piece. A hex nut was also fastened to the 3D printed piece  using two set screws to approximate the total payload during operating conditions. The testing setup is illustrated in Figure \ref{fig:shaker_test}3A. The time series of axial acceleration was then recorded as a function of frequency. Each data point was recorded by printing the  analog signal output to the serial port, which was then exported to a .txt file using the freeware CoolTerm. Since the  magnitude of acceleration, $\gamma$, is given by $\gamma=A(2\pi f)^2$, the amplitude of oscillation at each frequency can be calculated and is plotted in Figure \ref{fig:shaker_test}3B. The total harmonic distortion (THD), which characterizes the purity of the fundamental signal, was also measured at 30, 60, and 90 Hz at increasing amplitudes. No apparent trend in the THD value was observed across amplitude or frequency, with the mean value across all parameters found to be $12.7 \pm 6.9 \%$. {\color{black} The experimental data presented in this section is meant to be representative of all units using a 3'', 15 Watt speaker. Although not essential to test each unit, if users suspect their driver of having significantly different performance, the tests detailed here should be repeated to evaluate the operating specifications.}



\section{(3) Application}\label{h.f78bi3oom0mu}

\subsection{Use case(s)}\label{h.4q5g9edishy3}
The primary use case for this lab kit is to supplement the learning of undergraduate students studying mechanical vibration. In total, the vibration laboratory kit can be used to perform five different experiments. In order to reduce the cost of each kit and ensure simplicity of operation, no sensors are used to take quantitative measurements. However, by utilizing strobing to highlight the qualitative behavior, students can nonetheless observe important trends in a physical system, and compare these trends directly to theoretical predictions. Sample lab assignments for each experiment are included in the document repository. 

\subsection{One degree of freedom}

The first lab consists of investigating a simple forced one degree of freedom spring-mass system and is shown in Figure \ref{fig:1DOF}4. The system represents a base excitation problem where a payload is forced by a spring attached to a vibrating base. Here, the shaker is mounted face down on the top panel of the housing. A drive rod attaches a coupler to the shaker. Then, a spring is threaded to the coupler and a steel hex nut is attached, which acts as the primary payload. With the spring constant and mass of the nut provided, the students are asked to first predict the natural frequency of the one degree of freedom system with the single nut and then when a magnet of known mass is attached to the nut. For a forced spring-mass system, the natural frequency is given by $f_n=\frac{1}{2\pi}\sqrt{\frac{k}{m}}$ where $k$ is the spring constant and $m$ is the mass of the payload. Then, students are asked to perform an experimental frequency sweep utilizing the LED strobe lights to determine the frequency corresponding to the maximum amplitude amplification of the mass for these two configurations and compare their experimental observations with the theoretical prediction. The data in Figure 4B is presented in the form of a ``spatio-temporal'' diagram.  The goal of such graphical representations is to illustrate the time dynamics of a 1D system using a single image, derived directly from a video recording.  To create these images, narrow slices of pixels are extracted from a video and then stacked next to each other to form a single image. Thus rather than the axes of the image representing two spatial coordinates as in a standard photograph, one axis represents space while the other represents time.


\subsection{Passive vibration isolation}
The second lab is focused on investigating passive vibration isolation and is shown in Figure \ref{fig:isolation}5. Here, the shaker is removed from the housing and placed upright on a flat surface. A 3D printed assembly is mounted to the shaker via the threaded drive rod. The assembly consists of springs that are sandwiched between a top and bottom platform. However, by threading screws through the top and bottom of both platforms, the springs can be effectively locked in place, and the assembly vibrates as a rigid body. Students are asked to place dried beans on the surface of the top platform and observe the magnitude of the vibrations at a frequency well above the natural frequency of the spring-supported platform system.  When driving the shaker above a critical frequency (theoretically when $f>\sqrt{2}f_n$), students will notice that the amplitude of the beans is reduced when the springs are free to move (i.e. the payload is isolated) as compared to the rigidly connected case.  In contrast, when driving the shaker below a critical frequency (theoretically when $f<\sqrt{2}f_n$) the transmitted force to the top platform is amplified rather than isolated when using the passive spring isolator.  A key lesson of this laboratory is the importance of selecting isolator parameters carefully when designing a passive vibration isolation system.

\subsection{Nonlinear oscillator}
The third lab consists of investigating the dynamics of a nonlinear oscillator and is shown in Figure \ref{fig:nonlinear}6. This system consists a thin {\color{black} ferromagnetic} metal beam mounted over two permanent magnets equidistant from the beam's centerline and is a classical model system used in the study of nonlinear dynamics \citep{moon1979magnetoelastic}. Due to the induced-dipole attraction of the magnets on the beam, the beam can deflect towards either magnet. This model system can be described by the Duffing equation, where the magnetic force acts as a nonlinear spring. The approximate equation of motion for the tip position $x(t)$ of the beam is given by
\begin{equation*}
    m_e \ddot{x} + c \dot{x} + k_s x = F_0 \cos(\omega t) + F_m(x)
\end{equation*}

where $m_e$ is the {\color{black} effective} mass of the beam, $c$ is the structural damping coefficient, $k_s$ is the effective spring constant of the beam, $F_0$ is the forcing amplitude, and $\omega$ is the forcing frequency. The magnetic force is given (to leading order) by $F_m(x)=k_m [1-(\frac{x}{d})^2]x$ where $k_m \geq 0$ is a parameter that depends on the relative height between the magnets and the beam. With this equation, the students are asked to determine the equilibrium positions of this system and assess their stability as a function of the magnetic parameter $k_m$. Using stability analysis, it can be found that when $k_m > k_s$, the $x=0$ equilibrium position is unstable, while the $x=\pm d \sqrt{1-\frac{k_s}{k_m}}$ position is stable. When $k_m < k_s$, the $x=0$ equilibrium position is the only solution, and is stable. Since $k_m$ increases as the beam is brought closer to the magnets, the students can qualitatively test and observe the static bifurcation.  

In this setup, the top panel of the housing assembly is rotated 90 degrees (as compared to the configuration shown in Figure 1) and mounted on the front face of the housing. This configuration allows the shaker to be mounted with the drive axis horizontal. A thin, {\color{black} ferromagnetic} rectangular sheet metal beam is clamped to the drive rod on the shaker using nuts, which allows the beam to oscillate over each magnet. The base panel of the housing is shifted outwards such that it lies directly under the shaker, and another sheet metal plate is fastened to the surface of the base panel to allow the magnets to be positioned and held securely in place. Students are then asked to experimentally observe three regimes as the amplitude of the vibration is increased progressively: (1) small oscillations about one of the equilibrium positions, (2) chaotic/irregular oscillations between the two equilibrium positions, (3) large periodic oscillations passing over both equilibrium positions on each cycle. Students are additionally asked to numerically simulate Duffing’s equation to generate three phase portraits that correspond qualitatively to the three regimes observed in the experiment.  Students are also encouraged to adjust parameters in both the experiment and simulation in order to identify other possible vibration modes in this system.

\subsection{Multi-degree of freedom}
The fourth lab consists of investigating a simple two degree of freedom spring-mass system and is shown in Figure \ref{fig:2DOF}7. The physical configuration of the system is identical to the one degree of freedom system. Here, the setup is slightly modified by threading a second spring and mass (hex nut) to the base of the original mass. Before experimentally investigating this system, students are asked to first write down the governing equation for the motion of both masses, given by $x_1(t)$ and $x_2(t)$. For the purposes of the model, the system is assumed to be undamped and the equations of motion are thus given by
 \[
    \left\{
                \begin{array}{ll}
                  m \ddot{x_1} = k(x_2-x_1)+k(y-x_1)\\
                  m \ddot{x_2} = k(x_1-x_2)
                \end{array}
              \right.
  \]
where $y$ is the motion of the driver.  The students are asked to solve the governing equations in order to determine the {\it two} natural frequencies of the system as well as the corresponding mode shapes. 
Then, by once again performing an experimental frequency sweep, students experimentally verify their predictions with the aid of strobing to gauge the phase difference and relative vibration amplitudes. There is also an intermediate frequency at which the first mass is stationary while the second mass and driver oscillate out of phase. Students are also asked to predict and observe this state.

\subsection{Continuous system: One-dimensional wave}
The fifth lab consists of investigating the resonant modes of a tensioned string under harmonic forcing and is shown in Figure \ref{fig:wave}8. The lab allows the students to explore how the resonant frequencies are related to the string’s tension and length. Here, the shaker is once again mounted on the top panel of the housing. A string is attached to the shaker's drive rod, and an additional stand manufactured using laser-cut MDF is required so that the string can be tensioned by a known amount. Washers are hung vertically at the end of the string, from which the tension of the string can be estimated by the mass of the washers. 

The system can be modeled by assuming that the string of length $L$ has constant linear density $\rho$ and is under a constant tension $T$. Its transverse displacement $u(x,t)$ is governed by the linear PDE
\begin{equation*}
    \frac{\partial^2 u}{\partial t^2}=\frac{T}{\rho} \frac{\partial^2 u}{\partial x^2}.
\end{equation*}

In the experiment, the boundary conditions are that one end is shaken harmonically with amplitude $u(x = 0,t) = a\cos(\omega t)$, and the other end is held fixed such that $u(x = L,t) = 0$. Students are asked to solve this boundary value problem for the steady-state shape of the string and identify the resonant frequencies and mode shapes. 

In the experiment, students are asked to first measure out a string of a certain length as well as the total mass of the washers to determine the tension in the string. Then, by performing an experimental sweep over vibration frequencies, students visually identify the frequencies at which they experimentally observe the first three harmonic modes. The same test is repeated, but this time with the tension of the string changed by either adding or removing weights. Finally, the test is repeated for a third time, but this time keeping the tension of the string fixed, while changing the length of the string. Using the expressions derived in the first part of this problem, mainly that the natural frequencies are given by
\begin{equation*}
    f_n=\frac{1}{2\pi}\sqrt{\frac{T}{\rho}}\frac{n \pi}{L}, n=1,2,...,
\end{equation*}
the students are asked to predict the ratio of the original fundamental resonant frequency and the new fundamental resonant frequency (after either the tension or the string length has been changed). Then, the experimentally determined ratios are also calculated and compared to the theoretical predictions.



\subsection{Reuse potential and adaptability}\label{h.6wkumyl0ejrh}
The platform developed can be extended to demonstrate various additional vibration phenomena. For instance, using the existing setup, the nonlinear oscillator lab setup can easily be adapted to demonstrate the transverse vibration modes of a cantilevered beam and the associated deflection shapes corresponding to resonant frequencies. By extending the capabilities of the GUI platform used to drive the shaker, experiments could be done with more complex drive signals such as general periodic signals (e.g. concepts of Fourier series) or response to non-periodic forcing (e.g. unit impulse response). The current setup can be further extended to support additional vibration experiments with additional hardware. For example, by constructing an annular frame and wrapping an elastic sheet around it, the shaker can be used to vibrate a circular membrane and demonstrate the resonant vibration modes of a continuous system in two spatial dimensions.  





\section{(4) Build Details}\label{h.l8i9vokvs0bj}

\subsection{Availability of materials and methods}\label{h.60suejv0jlzi}
The housing assembly panels can be laser cut out of any 1/4" thick material, including acrylic or plywood. Components requiring 3D printing will show the best results using a SLA printer, but any commercially available FDM printer that uses filaments should be able to achieve the same results given that threads can be properly tapped by hand. Additional hardware components including fasteners, springs, and sheet metal have been sourced from McMaster-Carr {\color{black}(online supplier of hardware and raw materials based in the US, https://www.mcmaster.com/)} However, these components should also be readily accessible at most hardware stores. 

For the electronic assembly, only standard electronic components were sourced. However, it is recommended that a digital amplifier with at least two output channels be used so that the lighting and shaker can be driven independently. In terms of software, the user should have access to MATLAB to load the GUI used to drive the shaker and lighting.


\subsection{Ease of build}\label{h.wg823sgyb1e4}
The laboratory kit has been designed such that it can be easily assembled by undergraduate students. For the hardware, the housing assembly only requires fasteners to secure the individual panels without any need of permanent adhesives.  This feature allows it to be assembled and reconfigured with minimal effort.  For the electronics, no soldering is required with only minimal assembly using jumper wires. 

In order to manufacture the lab kits, knowledge of 3D printing and laser cutting is needed. Epoxy is used in one case to permanently secure the cylindrical piece to the voice coil of the speaker. A full set of manufacturing and assembly instructions as well as a bill of materials is included in the document repository.    

\subsection{Operating software and peripherals}\label{h.uz77dixfh5i4}
A GUI designed and operated in MATLAB is used to drive the shaker. The signal processing toolbox should also be added to MATLAB. All 3D printing files were designed in Autodesk Fusion 360, and the native f3d or exported step files can be directly edited using the same software or another CAD program. Laser cutting vector files were prepared and can be edited using Adobe Illustrator or an open-source alternative.  



\subsection{Dependencies}\label{h.vr0vnjs8z9ar}
The laboratory kit uses a GUI designed and operated in MATLAB.  As an alternative, we have also developed a dual-channel tone generator that can be run in Octave which accomplishes the same tasks.  However, in cases where MATLAB {\color{black} and Octave are} not accessible to the user, a standard sine-wave function generator can be used in place. Another option is to use a smartphone function generator app to drive the shaker and lighting as demonstrated in prior work \citep{harris2017visualization}. Specific electronic hardware is recommended in the bill of materials, but the laboratory kit is not dependent on these exact selections. Notably, nearly any dual-channel amplifier can be used in place of the recommended model. Furthermore, any similarly-sized (or larger) speaker can be used in place of the recommended model, however, the cylindrical insert that allows the drive rod to be interfaced must be edited appropriately to match its outer diameter to the inner diameter of the voice coil. 


\subsection{Hardware documentation and files location:}\label{h.nbisrsde6sc3}
{\color{black}
Archive for hardware documentation and build files.

Name: Vibrations-Demo

Persistent identifier: https://doi.org/10.5281/zenodo.5888554

License: Creative Commons Attribution 4.0 International (documentation), CERN Open Hardware Licence Version 2 (hardware and software)

Publisher: Daniel M. Harris

Date published: 21/01/2022

This archive contains the modifiable build files and software at the time of publication. The current version is available in a GitHub repository at https://github.com/harrislab-brown/Vibrations-Demo.
}












\section{(5) Discussion}\label{h.90jl7wm65t65}

\subsection{Conclusions}\label{h.h3fr33ylzsnh}
In this work, we've described a reconfigurable laboratory kit for use in an undergraduate mechanical vibrations course. In addition to the source files and documentation for the hardware, resources provided include five companion laboratory assignments that students can complete remotely.  

At the end of the Fall 2020 semester, all 15 students in the course completed University-administered course evaluations, which were supplemented by the instructor with specific questions regarding the laboratory kits.  In particular, students were asked to describe whether they agree or disagree with various statements by providing an integer score on a Likert scale of 1 (corresponding to ``Strongly Agree'') to 5 (corresponding to ``Strongly Disagree'').  Of particular note, all 15 students {\it strongly agreed} that the lab kits were effective.  Furthermore, 14/15 students {\it strongly agreed} that the kits should be used again in future years (with the remaining student providing a numerical score of `2 = Agree'). In open responses, individual students noted that the kits ``were easy to use and demonstrated the principles learned in class effectively'' and that the ``modularity of the assembly pieces was an ingenious way to pack multiple experiments worth of material into a single assembly''.  Overall, one student described the kit as a ``really unique way to engage with the course material in a hands-on way''.


\subsection{Future Work}\label{h.neocsr410zj}
The laboratory kit is designed in a reconfigurable way such that additional laboratory modules could be readily developed using the same basic hardware.  Future modules might focus on general periodic or aperiodic forcing, active vibration control, modal analysis, or random vibration.  The laboratory experiments described could also be readily made further quantitative by incorporating low-cost accelerometers and a data acquisition system (via Arduino, for instance), however, such enhancements would come with additional cost, as well as setup and operational complexity.  

Even outside of a remote learning setting, it is anticipated that the hands-on nature of the reconfigurable lab kit will benefit student learning.  We suspect that these kits allow students to} build a more holistic understanding of laboratory hardware and vibrations phenomena, when compared to a more traditional fully pre-prepared engineering laboratory.  Future work will aim to assess student learning outcomes by use of such a kit in an ``in-person'' learning environment. 

\subsection{Paper author contributions}\label{h.fy8hbipy6kwe}

IH designed and manufactured the hardware.  AHK developed the graphical user interface (GUI).  IH and DMH tested the device, developed the lab assignments, and applied them in the classroom setting.  IH and DMH wrote the manuscript and prepared the file repository.  All authors reviewed the manuscript prior to submission.


\subsection{Acknowledgements}\label{h.gu3yyarx72d6}

We gratefully acknowledge the students enrolled in ENGN 1735/2735: Vibrations of Mechanical Systems in Fall 2020 Brown University who tested the device throughout the semester and provided valuable user feedback.  In particular, we thank Jacob Morse for specific suggestions which led to improvements in the hardware design for the vibration isolation lab.  We also acknowledge the staff of the Brown Design Workshop for manufacturing support.



\subsection{Funding statement}\label{h.4u1a7tugh2om}

This work was funded by an HHMI grant (\#52008092, Program Director, Mark Johnson), Brown University's School of Engineering, and Curriculum Development Funds for Undergraduate STEM Courses administered through Brown University.


\subsection{Competing interests}\label{h.q1j1rznb43fl}

The authors declare that they have no competing interests.





\bibliography{bibliography_vibrations}

\section{Copyright notice}\label{h.jm5gcqv4g8x0}
\textcopyright 2022 The Author(s). This is an open-access article distributed under the terms of the Creative Commons Attribution 4.0 International License (CC-BY 4.0), which permits unrestricted use, distribution, and reproduction in any medium, provided the original author and source are credited. See http://creativecommons.org/ licenses/by/4.0/.




\newpage
    
\newpage
    
\newpage

\newpage
    \begin{figure}[hbt!]
    {\centering
    \includegraphics[width=350pt]{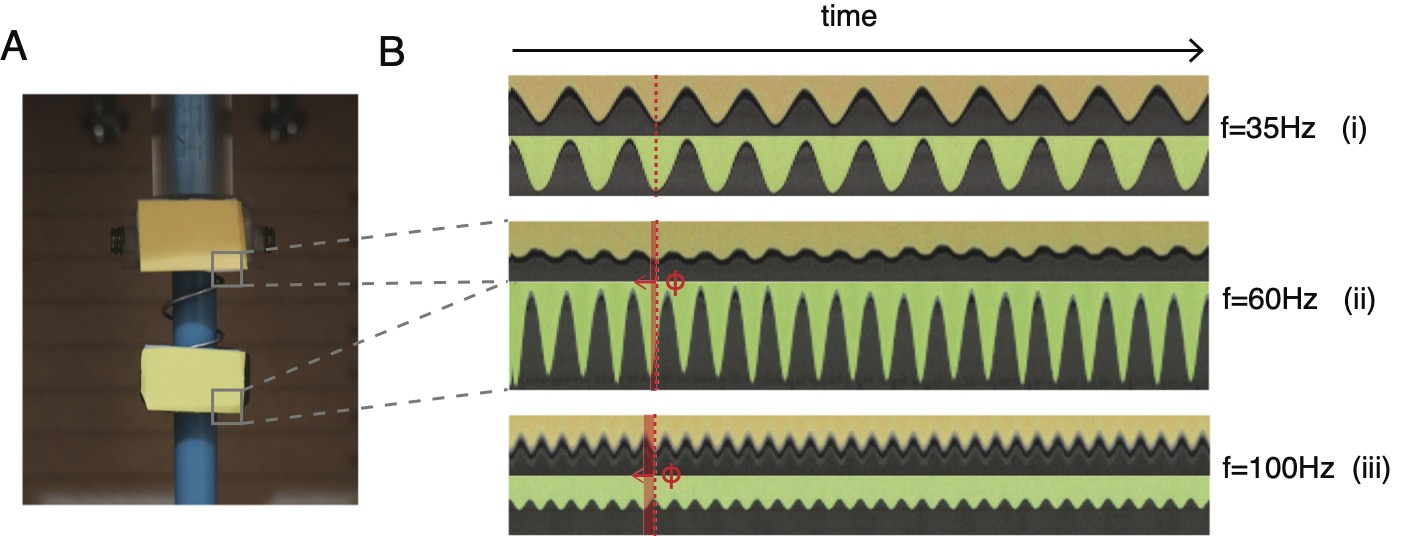}
    \caption{\small One degree of freedom lab. (A) Close-up image of the experiment setup consisting of a hex nut suspended on a spring, and connected to the driver. (B) Spatio-temporal diagrams of three different vibration regimes. In each case, the top (orange) waveform depicts the driving oscillation, while the bottom (yellow) waveform depicts the vibrating mass. (i) Below resonance (35 Hz): the amplitudes are approximately equal, and oscillation is in phase with the driving. (ii) Near resonance (60 Hz): the response amplitude of the mass is much greater than that of the driving with a phase difference of approximately 90 degrees. (iii) Above resonance (100 Hz): the response amplitude of the mass is less than that of the driving with a phase difference of approximately 180 degrees.}
    }
    \label{fig:1DOF}
\end{figure} 
    
\newpage
    \begin{figure}[hbt!]
    {\centering
    \includegraphics[width=350pt]{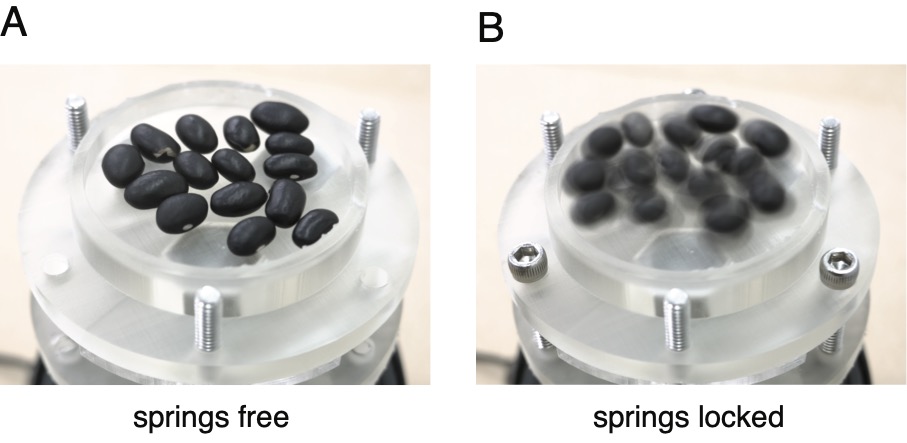}
    \caption{\small Passive vibration isolation lab.  Dried beans are placed on a passive isolation device and vibrated at 120 Hz. (A) The support springs are free to oscillate and successfully isolate the top platform from vibrations. (B) Under the same vibration conditions, the springs are locked, and the vibration is transmitted to the top platform causing the beans to jump about the tray.  Both images are created by overlaying several successive frames from video recordings.}
    }
    \label{fig:isolation}
\end{figure} 

\newpage
    \begin{figure}[hbt!]
    {\centering
    \includegraphics[width=350pt]{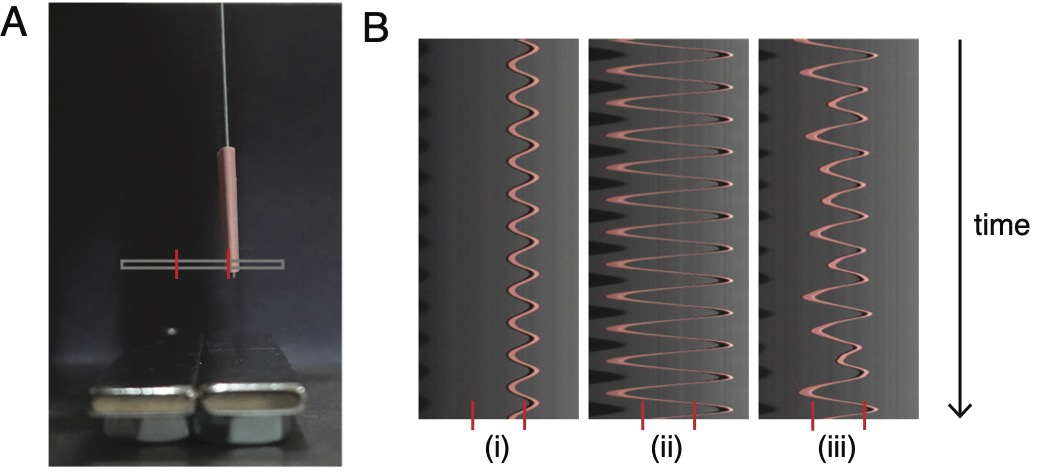}
    \caption{\small Nonlinear vibration lab. (A) Image of the experiment setup. The vertical red lines represent the approximate equilibrium positions. The horizontal gray bar represents the section from which the spatio-temporal diagrams are constructed. B) Spatio-temporal diagrams for (i) small amplitude oscillation about one fixed point, (ii) limit cycle oscillation over both fixed points, and (iii) chaotic/aperiodic oscillation.  The beam is being driven at 20 Hz with different amplitudes in each of these examples.}
    }
    \label{fig:nonlinear}
\end{figure} 
    
\newpage
    \begin{figure}[hbt!]
    {\centering
    \includegraphics[width=350pt]{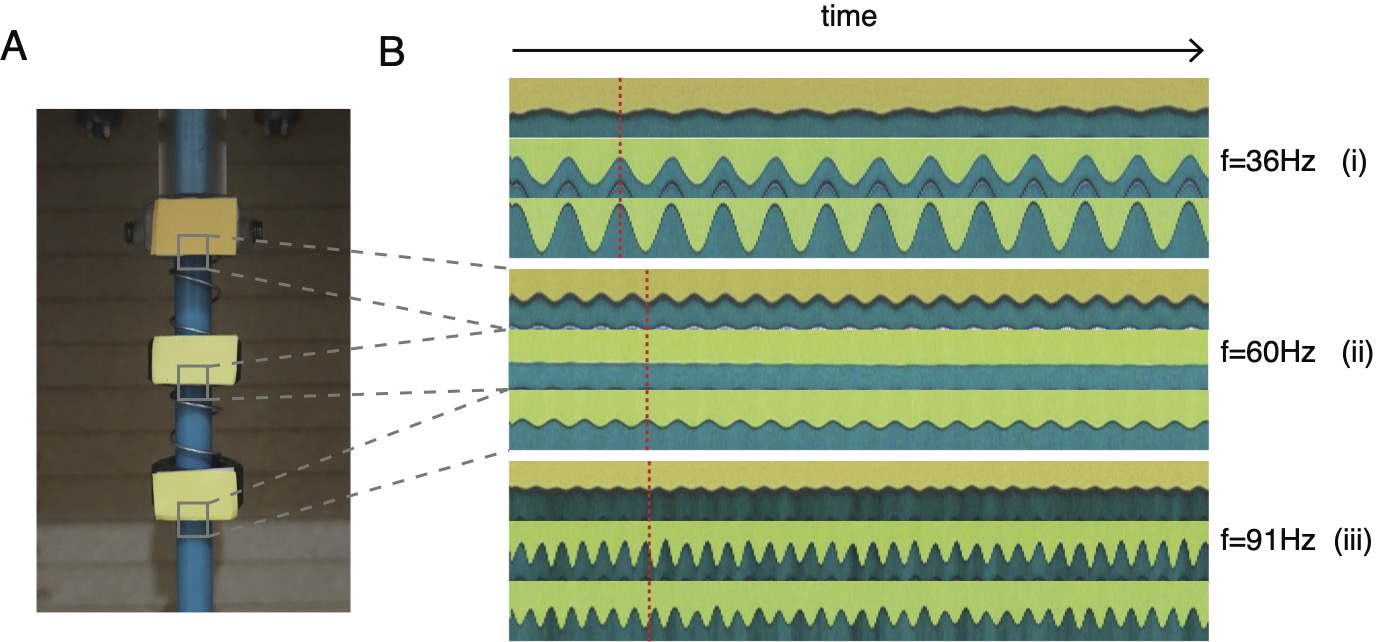}
    \caption{\small Two degree of freedom lab. (A) Image of the experiment setup. (B) Spatio-temporal diagrams of three different behaviors.  The top (orange) waveform represents the driving oscillation, while the middle and bottom (yellow) waveforms represent the vibrating masses. (i) The first resonant frequency at 36 Hz. The oscillation of both masses are in phase. (ii) Oscillation at 60 Hz: the middle mass is nearly stationary while the lower mass moves out of phase with the driving. (iii) The second resonant frequency at 91 Hz.}
    }
    \label{fig:2DOF}
\end{figure} 
    
\newpage
    \begin{figure}[hbt!]
    {\centering
    \includegraphics[width=350pt]{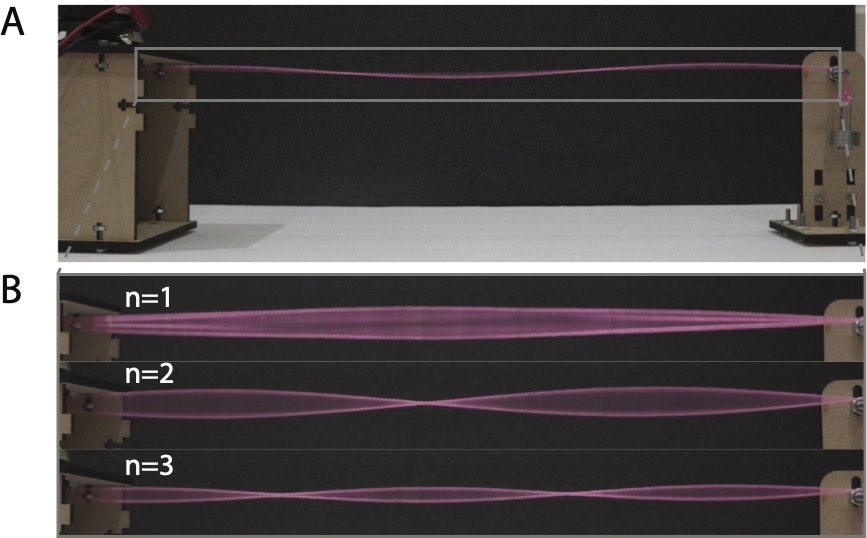}
    \caption{\small Continuous system (one-dimensional wave) lab. (A) Image of the experimental setup. (B) Images of the first three harmonics are visualized by overlaying successive frames from video recordings. The strings were vibrated at $f$=15, 30, and 45 Hz in order to observe modes $n$=1, 2, and 3 respectively. The tension on the string was fixed by a suspended weight of 30.1 g ( corresponding to a tension of 0.295 N).}
    }
    \label{fig:wave}
\end{figure}

\end{document}